\documentclass[aps,prd,amsmath,amssymb,12pt]{revtex4}
\usepackage{graphicx}
\usepackage{bm}
\def\d{{\rm d}}

\def\LL#1{\Lambda_{{#1}{\rm L}}}
\def\rl#1{\Lambda_{{#1}{\rm R}}}

\def\BRA{\left\langle}
\def\KET{\right\rangle}

\def\LK{\left(}
\def\RK{\right)}
\def\LBK{\left\lbrack}
\def\RBK{\right\rbrack}
\def\LB{\left\lbrace}

\def\RB{\right\rbrace}

\def\Eq#1{(\ref{#1})}

\def\sp#1#2{\left( #1 \vert  #2 \right)}
\def\lv#1{\left( #1 \right\vert}
\def\rv#1{\left\vert  #1 \right)}

\def\LMID{\left.\mid}
\def\RMID{\mid\right.}
\def\det{{\rm det}\,}

\def\DL{{\cal L}}
\def\DQ{{\cal Q}}
\def\DP{{\cal P}}

\def\Tr{{\rm Tr\, }}

\def\be{\begin{equation}}
\def\ee{\end{equation}}
\def\bea{\begin{eqnarray}}
\def\eea{\end{eqnarray}}
\def\bef{\begin{figure}}
\def\ef{\end{figure}}

\def\dt{{\rm d}t}
\def\emph#1{{\bf{#1}}}

\def\Tr{\mathop{\rm Tr}}

\def\L{{\cal L}}
\def\journal#1#2#3#4{{#1} {\bf #2}, #3 (#4)}

\begin{document}
\title{On  Generated Dynamics for Open Quantum Systems: Spectral Analysis of Effective Liouville}
\author{Martin Jan{\ss}en}
\email{mj@thp.uni-koeln.de}
\affiliation{Institut~f\"ur~Theoretische~Physik, Universit\"at~zu~K\"oln\\ Z\"ulpicher~Str.~77, D-50937~K\"oln}
\date{July 30, 2017}
\begin{abstract}
We point out that the quantum dynamical map of an open quantum system can be generated  by an effective  Liouville operator.
The effective Liouville shows the dynamical breaking of time reversibility. This breaking of reversibility is  expressed by the effective Liouville's discrete spectrum having negative imaginary parts. This generated dynamics of open quantum systems is capable of memory effects described by a frequency dependence of the  spectrum. 
When memory effects can be neglected or smoothed out, the effective Liouville  generates the well known  semi-group dynamics of open systems in the Markov approximation.
The spectral analysis of the effective Liouville -with or without memory- allows to represent the quantum dynamical map and  expectation values of physical quantities by metric expressions using complex eigenvalues and bi-orthonormal eigenmodes. The long time dynamics is dominated by the zero-mode, which forms the  stationary state of the open system. The relaxation time scale $\tau$  is set by the inverse of the smallest negative imaginary part  of the non-vanishing effective eigenvalues of the effective Liouville.  In generic cases of system-environment-coupling this time scale $\tau$ is the relevant scale for both, the relaxation of diagonal elements (populations) and the decay of off-diagonal elements (coherences) of an initial density matrix in the stationary state's diagonal representation. It  also sets the time scale after which the  entropy  tends to its stationary value. The effective Liouville and its spectral content thus give a framework to study the dynamic  breaking of time reversibility, memory effects, decoherence, relaxation to a stationary state and the role of entropy in open quantum systems.
\end{abstract}
\maketitle

\section{Introduction}
A system coupled to an environment  forming a total closed quantum system with a von Neumann dynamics is the common framework to study  open quantum systems (see ~\cite{BreuPet},~\cite{MarPuet} for an introduction).  If one  neglects the entanglement of the density operator between system and environment at some single moment, one can use this as a separated initial condition to gain a closed dynamic description for the density operator reduced to the system. This dynamic is known (see e.g.~Sec.~II A.1 in \cite{BreuerEtal}) to be represented as  a completely positive trace preserving map, denoted as  quantum dynamical map. It is also known that every quantum dynamical map has a so called Kraus decomposition  which guarantees that the system's density operator $\rho(t)$  keeps to be a  density operator during its time evolution. However, the decomposition is not unique and not defined in a constructive way when starting from microscopic Hamiltonians. Under the further assumption of the dynamic map having the semi-group property (the so called Markov approximation), a time-independent generator
 $\L$ for the dynamic exists,  
\be
\dot{\rho}(t)= -i \L  \rho(t) \, . \label{1.1}
\ee
This generator $-i\L$ can be written in the so called Lindblad form which then guarantees  that the system's density operator $\rho(t)$  keeps to be a  density operator during its time evolution. 
Again, the Lindblad form is not unique and not defined in a constructive way starting from microscopic Hamiltonians. Under further approximations it can be made constructive starting from microscopic Hamiltonians. The approximations  usually rely on  single scattering interactions or on weak interactions (or on both) (for discussions see e.g. Sec.~5.4 in \cite{Horn}). The Markov approximation can be relaxed when it is possible to invert the dynamic map at each time $t$. Then a time local version of \Eq{1.1} can be written down with a time dependent generator $\L(t)$, for which a similar to Lindblad representation can be written (see e.g. Eq.~(13) in \cite{BreuerEtal}). Again, the representation is not constructive starting from microscopic Hamiltonians, except for interactions via single particle-particle exchanges (see e.g. \cite{ZhangEtal}). It can, however, be used  as a constructive intermediate model in a jump modeling approach (see e.g. \cite{PiiloEtal}).

An  alternative  approach to open quantum systems is based on a microscopic Hamiltonian and on  reducing to the system by path integration in a closed time path  formalism  (see e.g.~\cite{CalzHu}).  
So far, it has been worked out for continuous variables and field variables in a perturbative way, but it is -in principle- capable of incorporating renormalization methods to go beyond perturbative results. A discussion of the limitations of the perturbative approach and of the single scattering approximation can be found in ~\cite{JPol_statdyn}.

It is also known since the early works of Nakajima \cite{Naka} and Zwanzig \cite{Zwan} that a quantum dynamical map can be written in a time convolutional form with a memory kernel $\L(t-t')$
\be
\dot{\rho}(t)= -i \int_{t_0}^t \dt' \L(t-t')  \rho(t') \,. \label{1.2}
\ee
To the authors knowledge this Nakajima-Zwanzig form of a quantum dynamical map has not been exploited very much beyond the weak coupling approximation and  it is stated often that its construction relies on the solution of the total system's dynamic.
In \cite{JanGen} the present author has pointed out that it is not necessary to solve the total system's dynamic to exploit the Nakajima-Zwanzig form. The key ist to use a Laplace transformed version of Eq.~({\ref{1.2}). An early attempt along the Laplace transformed version was undertaken by Fano in \cite{Fano}, however with a subsequent perturbative treatment. Here we follow a similar line with a non-perturbative spectral analysis to clarify the overall structure of possible time evolution in open quantum systems. 
 By using a Laplace transformed version of Eq.~(\ref{1.2}) and a well known algebraic method for constructing effective generators on a projected space it is possible to construct a generator for the system's reduced density matrix in frequency space,
\be
	\rho(z) = i \LBK z-\DL(z) \RBK^{-1} \rho_0 \, , \label{1.3}
\ee
where $\rho_0$ is the system's initial density operator and the effective Liouville $\DL(z)$ is a super operator acting only on the reduced system. Equation~(\ref{1.3}) is a powerful   substitute for Eq.~(\ref{1.1}) when a semi-group structure is missing due to memory effects (\ref{1.2}). Its construction can be made explicit in terms of projections to the system of the  total {Liouville operator} $\DL\cdot = \LBK H, \cdot\RBK$  and a separated  propagator  within an orthogonol complement to that projection. When the isolated system has a $d$-dimensional Hilbert space, the corresponding Hilbert-Schmidt space of trace class operators is made of $d^2$-dimensional matrices and the super operator acting on these matrices has dimension $d^2\times d^2$. Once it has been set up, the inversion of a $d^2\times d^2$-dimensional super matrix hast to be performed. To get the time dependent quantum dynamical map a Laplace back transform has to be added. Such situation also calls for a spectral analysis of the finite dimensional super operator to represent the quantum dynamical map of the density matrix and the expectation values of physical quantities by metric expressions using complex eigenvalues and bi-orthonormal eigenmodes.   These steps are described  in this comment. Related works on spectral analysis for quantum dynamical maps can be found  in \cite{Petrosky} and in \cite{BruzdaEtAL}. 

Based on these steps  of spectral analysis of the effective Liouville ${\cal L} (z)$   we collect a number of statements about open quantum systems with quantum dynamic maps.
\begin{enumerate}
\item Time reversibility of closed systems is  unstable with respect to environmental contact provided the system's energy spectrum can be considered as  discrete while the environment's spectrum  can be treated as smooth continuous.
\item The   time evolution of the system's density matrix is a sum of exponentially damped oscillations, relaxing to a stationary density matrix. This stationary state  is independent of the initial condition. The frequencies and relaxation rates of this motion are given by real parts and negative  imaginary parts, respectively,  of the solutions $z_k$ oft the equation $z=\lambda_k(z)$, where $\lambda_k(z)$ is the $k$-th eigenvalue of the effective Liouville ${\cal L} (z)$. 
 Accordingly, expectation values of observables  relax  to their stationary values  showing a superposition of exponentially damped oscillations.
\item The stationary density matrix $\rho_\infty$ is the normalized right zero mode of the effective Liouville which always exists due to probability conservation. It is unique as long as the system is not separated into independent subsystems.
\item The inverse oft the smallest negative imaginary part of the effective eigenvalues $z_k$ sets the global relaxation time scale $\tau$  for  the density matrix $\rho(t)$ to reach the stationary density matrix $\rho_\infty$. In the eigenbasis of  the stationary density matrix, under  conditions to be discussed in the text, $\tau$ is the decay time of off-diagonal elements (coherences)  and is also the relaxation time scale for  diagonal elements (populations) to reach the stationary distribution.
\item Density matrix elements in the eigenbasis of some other observable can only turn diagonal in the stationary limit to a precision given by the commutator of the observable with the stationary density matrix.
\item Pure dephasing  environments (coherences decay while populations stay constant) are non-generic, because the initial density matrix already has to have  stationary populations. This can only happen in the eigenbasis of some observable which expectation value keeps to be unaffected by  the presence of the environment.  For example it could happen in the energy eigenbasis when the system Hamiltonian commutes with the interaction Hamiltonian and there is exactly no energy exchange between system and environment - a clearly non-generic situation. 
\item In a generic situation, where some relaxation of populations occurs,  coherences typically decay with the same characteristic time scale of population relaxation. 
\item To have a  quick decay of coherences,   accompanied by a much slower  relaxation of populations  in some macroscopic modes,  the  separated scales    stem from two different ways of opening the closed system: a fast relaxation process with $\tau_1$ with little change of populations (e.g. almost  elastic scattering with external modes) and an additional slow relaxation process with $\tau_2\gg \tau_1$ (e.g. the few slow macroscopic modes of a system are coupled inelastically to  many fast irrelevant modes). After $\tau_1$, the  coherences  in the representation of the final stationary state may have shrunk already to small quasi-stationary values before they finaly, after time $\tau_2$, decay to zero. The relaxation process after time $\tau_1$ can then very well be modeled by an ordinary Markov process for only the populations.
\item The entropy of the system's density matrix tends  to its stationary value after the time $\tau$.
  For $t\gtrsim \tau$, in an overdamped system (or approximately in a low frequency underdamped system),
 the relative entropy production stays non-positive  and the relative entropy, being a non-negative function,  can serve as a Lyapunov function for the stability of the stationary state. This is not possible in an underdamped situation with essential oscillation. At time scales well  below 
$\tau$ the relative entropy can never serve as a Lyapunov function  because of irregular oscillations. 
\end{enumerate}
\section{The effective Liouville}
In constructing the reduced density operator of a system coupled to an environment 
we follow the quite general projector formalism. 
The  reduced density operator \index{relevant density operator} is defined by a projector $\DP$ on the Hilbert-Schmidt space of linear operators over the Hilbert space of the total system,
\be
\rho:= \DP \varrho \, .\label{4.50}
\ee
Here $\varrho$ is the total density operator of the total closed system containing relevant (system) and irrelevant (environment) variables. 
The standard example of a convenient projector $\DP$ can be defined via  a partial trace over the Hilbert space  of irrelevant (environmental) variables (denoted by subscript$E$),
\be
\DP \cdot = \Tr_{E} (\cdot) \otimes \rho_E \, , 
\ee
where $\rho_E$ is the would-be stationary density operator of the environment -  if the environment was not coupled to our system.
The projector on the complement to the relevant variables is denoted as $\DQ:=1-\DP$. A projector has the property that its square equals it.

The phrase  environment is not restricted to spatially external variables interacting by  scattering processes with  the system but may also mean variables spatially  within the system which  will (or cannot) be treated as relevant.
For example, few macroscopic slow hydrodynamic modes in a  fluid may be of relevance to follow and all other system variables still interacting in a complicated way  with the relevant modes are denoted  as irrelevant variables.  Thus, the configurations of the system are described  by the relevant variables  and the configurations of the total system by  relevant and irrelevant variables.

The dynamics of the total density operator is described by  a von Neumann equation  with total Hamiltonian $H_{\rm tot}$,
\be
\partial_t \varrho(t) = -i\DL_{\rm tot} \varrho(t) \label{4.52}\, ,
\ee
with the  {Liouville operator} $\DL_{\rm tot}\cdot = \LBK H_{\rm tot}, \cdot\RBK$ as the generator of the closed quantum dynamics. As a super operator it is hermitian and has a  real valued spectrum $\omega_{\alpha \beta}=\epsilon_\alpha-\epsilon_\beta$, where $\epsilon_\alpha$ are the energy eigenvalues  of the total Hamiltonian.

The formal solution of the closed quantum dynamics is the time evolution of the total density operator,
\be
	\varrho(t) = e^{-i\DL_{\rm tot} t} \varrho(0) \, .\label{4.53}
\ee
We like to construct the dynamic equation for the relevant density operator $\rho(t)$ by using the decomposition,
\be
\varrho =\rho+ \DQ \varrho \, . \label{4.51}
\ee
We use a concept known from a similar problem for the dynamics of quantum states, where  a so-called effective Hamiltonian is constructed. The basic  idea goes back to \cite{WeisWig}. For the projection
formalism see \cite{Fesh}. To use this concept  is possible by considering the spectral content of the  total time evolution operator $e^{-i\DL_{\rm tot} t}$. It is captured in the frequency dependent {resolvent}\index{resolvent} (-i times the Laplace transformed of the time evolution operator)
\be
 	  \LBK z- \DL_{\rm tot} \RBK^{-1} =-i \int_0^\infty {\d}t \, e^{izt} e^{-i\DL_{\rm tot} t} \, .\label{44.54}
\ee
The resolvent  is  analytic in the upper plane $z=\omega + i\epsilon$ and singular on the real valued spectrum of eigenvalues $\omega_{\alpha \beta}$. The dyads $\LMID \epsilon_\alpha \KET\BRA \epsilon_\beta \RMID $ are the corresponding eigenvectors of  $\DL_{\rm tot}$. Note that the eigenvalue $0$ is at least $d_{\rm tot}$-fold degenerate ($d_{\rm tot}$ being the dimension of the total system's Hilbert space)  for the total Liouville because of the commutator structure (even for non-degenerate energy eigenvalues). This fact   corresponds to  many possible choices for stationary density operators in the total system   $\varrho_{\rm st} = \sum_{\alpha=1}^{d_{\rm tot}} p_\alpha \LMID \epsilon_\alpha \KET\BRA \epsilon_\alpha \RMID $ with arbitrary normalized probability  distribution $p_\alpha$.

Now, the evolution of the total  density operator reads in frequency space
\be
	\varrho(z) = i \LBK z-\DL_{\rm tot} \RBK^{-1} \varrho_0 \, .\label{44.55}
\ee
For the relevant density operator a similar equation (see below \Eq{4.56}) holds for purely algebraic reasons, provided  we can treat the full density operator at some initial time as decoupled between system and environment,
\be
\varrho (t=0)=\rho_0 \otimes \rho_E \, .
\ee
This assumption is crucial  to reach a closed effective equation for the relevant density operator. It is well known that this assumption allows to represent the dynamics by a quantum dynamical map (see e.g.~\cite{BreuPet}). We will see in our treatment,  that the density operator  tends to  a stationary state, which is independent of the initial condition $\rho_0$. This shows that the initial condition is not  essential for the system's time evolution in the long time limit. When the system can never  be treated as decoupled from the environment,  the whole approach of finding a dynamical map  for the reduced  system  is misleading. In the rest of this note, we  stick to the dynamical map approach and consider its construction by a generator $\DL(z)$,
\be
	\rho(z) = i \LBK z-\DL(z) \RBK^{-1} \rho_0 \, .\label{4.56}
\ee
Here the effective Liouville operator \index{effective Liouville operator} $\DL(z)$ lives on the $\DP$-space only, but depends on the frequency. 
The effective Liouville  reads (for a short derivation see \cite{JanGen}) 
\be
	\fbox{$\displaystyle
\DL(z) =  \DL_{\DP} + \DL_{\DP\DQ} \LBK z- \DL_\DQ \RBK^{-1} \DL_{\DQ\DP}$}\, .\label{4.57}
\ee
The structure is formally the same as for effective Hamiltonians (see e.g. \cite{VolZel}). The terms have the following interpretation. The first term is the bare
Liouville in $\DP$-space. It is the system's Liouville in the absence of any couplings to irrelevant variables.  
The second term is
due to { virtual processes} in $\DP$-space. They are produced by
{hops} from $\DP$-space to $\DQ$-space, there taking a lift  with the
propagator\index{propagator} $[z-\DL_\DQ]^{-1}$ in $\DQ$-space, followed by hoping back
onto $\DP$-space (see Fig.~\ref{Fig4.1}).
\begin{figure}[t]
\centering
\includegraphics[width=10cm]{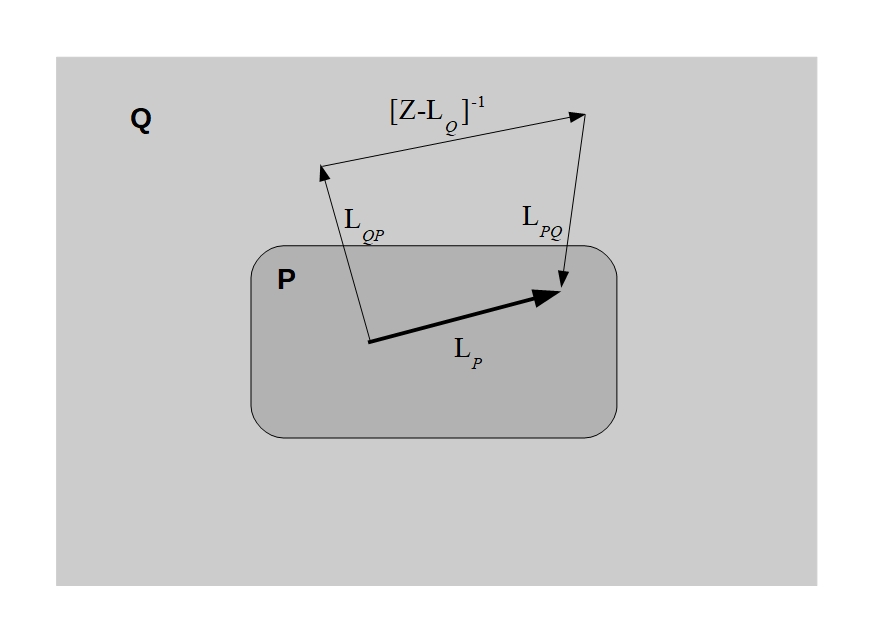}
\caption{Visualizing the effective Liouville in $\DP$-space by direct processes and virtual processes via propagation in $\DQ$-space}
\label{Fig4.1}
\end{figure}
Equation~(\ref{4.56}) is a quantum Master equation in frequency \index{quantum Master equation} space.
The effective Liouville is, in general,  no longer hermitian.  One can see this in the  decomposition  for $\epsilon \to 0$ of the resolvent in $ \DQ$-space  to separate real and imaginary parts,
\be
 \lim_{\epsilon \to 0} \LBK z-\DL_\DQ\RBK^{-1} = {\rm P} \LBK \omega -\DL_\DQ \RBK^{-1}  -   i\pi \delta (\omega -\DL_\DQ) \, ,\label{4.57a}
\ee
where $ {\rm  P}$ stands for the Cauchy value on integration. Integration over a dense spectrum of modes in $\DQ$-space resulting in a smooth  z-dependent anti-hermitian contribution 
$ -   i\pi \DL_{\DP\DQ} \delta (\omega -\DL_\DQ) \DL_{\DQ\DP}$  to the effective Liouville. This term breaks the hermiticity of the Liouville and thus the time reversibility of the effective dynamics for the open system.  Time reversibility of closed systems is therefore unstable with respect to environmental contact provided the system's energy spectrum can be considered as  discrete while the environment's spectrum  can be treated as smooth continuous -  the first statement of the introduction.

In the time dependent picture the quantum Master equation \Eq{4.56} has the Nakajima-Zwanzig form (see also Sec.~2.4.1~in~\cite{ZubEal} for a derivation)
\be
\partial_t \rho(t ) = -i\DL_\DP \rho(t) + \int_0^t \d\tau \,   \DL_{\DP\DQ}  e^{-i\DL_\DQ (t-\tau)} \DL_{\DQ\DP}\,  \rho(\tau) \, ,\label{4.460}
\ee
which shows a memory effect: all times between initial time $0$ and time $t$ contribute from the $\DQ$-journey.
This memory effect is mapped to frequency space by the Laplace transform to the frequency dependence of the virtual  process part of the effective Liouville. If the frequency dependence of the effective Liouville can be neglected or smoothed out, the effective Liouville is the semi-group generator of the quantum dynamical semi-group dynamics, as written in \Eq{1.1} (Markov approximation).
However, when this frequency dependence  cannot be neglected or smoothed out, the effective dynamics does not have the semi-group property in the time dependent picture and lots of solving techniques that rely on the semi-group property cannot be applied in the time dependent picture. The Laplace transformation  circumvents this problem: in frequency space the algebraic solving techniques are the same with or without memory. Once  the density operator $\rho(z)$ has been found  one may go back to the time dependence by inverse Laplace transformation. It is  not  necessary to perform the inverse Laplace transformation  if one only likes to study the long time behavior (or the short time behavior).  The long time behavior in an open quantum system is of main interest and the effective Liouville in frequency space is most convenient to address this long time behavior. 
 
So far, the closed algebraic equation for the  relevant  density operator represents the dynamics as  a matrix problem of multiplication and of inversion of given matrices -  perhaps assisted  by spectral analysis. The effective Liouville can be  constructed explicitly, once the system model Hamiltonian is defined and a  convenient propagator  $\LBK z-\DL_\DQ\RBK^{-1}$ has been chosen together with a  coupling of this propagator to the system by projection.
The corresponding matrix representation  will  have dimension $d^2\times d^2$ when the reduced system has a $d$-dimensional Hilbert space.  Thus, few relevant system variables and corresponding  states can  very well be be studied (e.g.~numerically) without further approximations. Technical details about matrix representations of super operators can be found in \cite{NamNak}

In the following we look at the spectral analysis of  Eq.~(\ref{4.56}).
\section{Solving the Dynamics By Spectral Analysis}
For the spectral analysis of non-hermitian super operators one uses a metric on the space of operators on which the effective Liouville might act. This is done by the Hilbert-Schmidt scalar product (on which the Hilbert-Schmidt norm is based) on the Hilbert-Schmidt space of operators as
\be 
	\sp{A}{B}:=\Tr \LB A^\dagger B\RB \, .
\ee
In the non-hermitian case a bi-orthogonal eigenbasis of right and left eigenvectors can be found 
\be
	{\cal L} (z) \rv{\rl{k}(z)} = \lambda_k(z) \rv{\rl{k}(z)}  \, ,
\ee
\be
	\lv{\LL{k}(z)} {\cal L} (z)  = \lambda_k(z) \lv{\LL{k}(z)} 
\ee
and, by proper normalization between left and right eigenvectors  $\sp{\LL{k}(z)}{\rl{k}(z)}=1$,   a decomposition of the super unity as
\be
	1= \sum_{k=0}^N   {\rv{\rl{k}(z)}\lv{\LL{k}(z)}}\, .
\ee
The number $N=d^2-1$ for a $d$-dimensional Hilbert space with associated $d^2$-dimensional Hilbert-Schmidt space.

Since the probability conservation
\be
\sp{1}{\rho(t)}=1
\ee
translates by Laplace transformation to
\be
\sp{1}{\rho(z)}=\frac{i}{z}\, ,
\ee
the effective Liouville must satisfy
\be
\lv{1}{\cal L}(z)=0\, .
\ee
Thus, the unit matrix $1$ is always a left eigenvector to ${\cal L}(z)$ with eigenvalue $0$.
By transposing the secular equation for this eigenvalue a right eigenvector with eigenvalue $0$ must exist, too.
We call this right eigenvector the zero-mode $\rv{0}$ and let it correspond to the index $k=0$. Since the eigenvectors of the effective Liouville are matrices we will also call them eigenmatrices.
Since the eigenvalue equation for right eigenmatrices operated from the left with the unit matrix yields
\be
	\sp{1}{{\cal L}(z) \rl{k}(z)}=\sp{1}{\rl{k}(z)} \lambda_k(z) =0
\ee
we can conclude: While for the zero-moder a finite value of $\sp{1}{0}\neq 0$ results, all other right eigenmatrices with non-vanishing eigenvalues have vanishing scalar product with the unit matrix (they are traceless matrices),
\be
\sp{1}{\rl{k}(z)}= 0 = \Tr \rl{k}(z) \; {\rm for}\; k\neq 0 \, .
\ee
The normalized zero-mode  is the stationary density matrix denoted as
\be
\rv{\rho_\infty} := \rv{0} \, .
\ee
The equation of motion in frequency space reads in spectral representation
\be
              \rho(z) = \sum_{k=0}^N    {\sp{\LL{k}(z)}{\rho_0}}
	\frac{i}{z-\lambda_k(z)} \rl{k}(z) \, .
\ee
We assume that the eigenmatrices and eigenvalues do not show a singular behavior with respect to the variable $z$. This will be the case when  the spectrum of the environmental Liouville alone has  a structureless spectrum (dense and gapless, no edges or singular states). Under such conditions the only singular points of $\rho(z)$ in the lower complex plane stem from the poles produced by $\frac{i}{z-\lambda_k(z)}$.  It is important to realize  that the poles can only lie in the lower $z$-plane, as required by the analytic properties of the full resolvent and further substantiated by the positive  character of  the term  $ \DL_{\DP\DQ} \delta (\omega -\DL_\DQ) \DL_{\DQ\DP}$ appearing with $-i\pi$ as a prefactor in the effective Liouville. The negative imaginary parts show the instability of time reversibility against environmental contact (for a related discussion see \cite{JPol_arrow}).  For a sketch of the spectral properties of isolated and open systems see Fig.~\ref{Fig5.1}. This is our detailed argument in favor of statement 1. of the introduction.
In the  Markov approximation the $z$-dependence of eigenmatrices and eigenvalues is absent at all and poles are only at $z=\lambda_k$ in the lower $z-$plane. In the non-Markovian non-singular case we  thus have an analytic  z-dependence such that $\lambda_k(z)$ form frequency bands ($z=\omega+i0$).
The Laplace back transform can then be calculated with the help of residue and is found to be
\be
            \rho(t) = \rho_\infty + \sum_{k=1}^N    a_k 
	{e^{-iz_kt}} \rl{k} \, , \label{last}
\ee
where the solutions $z_k=\omega_k-i\gamma_k$ with $\gamma_k > 0$ of 
\be
z=\lambda_k(z)
\ee
will be called effective eigenvalues of the effective Liouville. The complex prefactors $a_k$ are given as
\be
a_k = \frac{ {\sp{\LL{k}(z_k)}{\rho_0}}}{1-\lambda_k'}
\ee
where $\lambda_k'$ is the first derivative of  $\lambda_k(z)$ at value $z=z_k$. Note that we assumed non-degenerate effective eigenvalues and non-crossing bands $\lambda_k(z)$. Degenerate eigenvalues are non-generic and  may happen accidentally resulting in additional terms with time dependence of type $t^{n}e^{-iz_kt}$ with $n=1,2, \ldots M$ for a $M$-fold degenerate eigenvalue.

\begin{figure}[t]
\centering
\includegraphics[width=10cm]{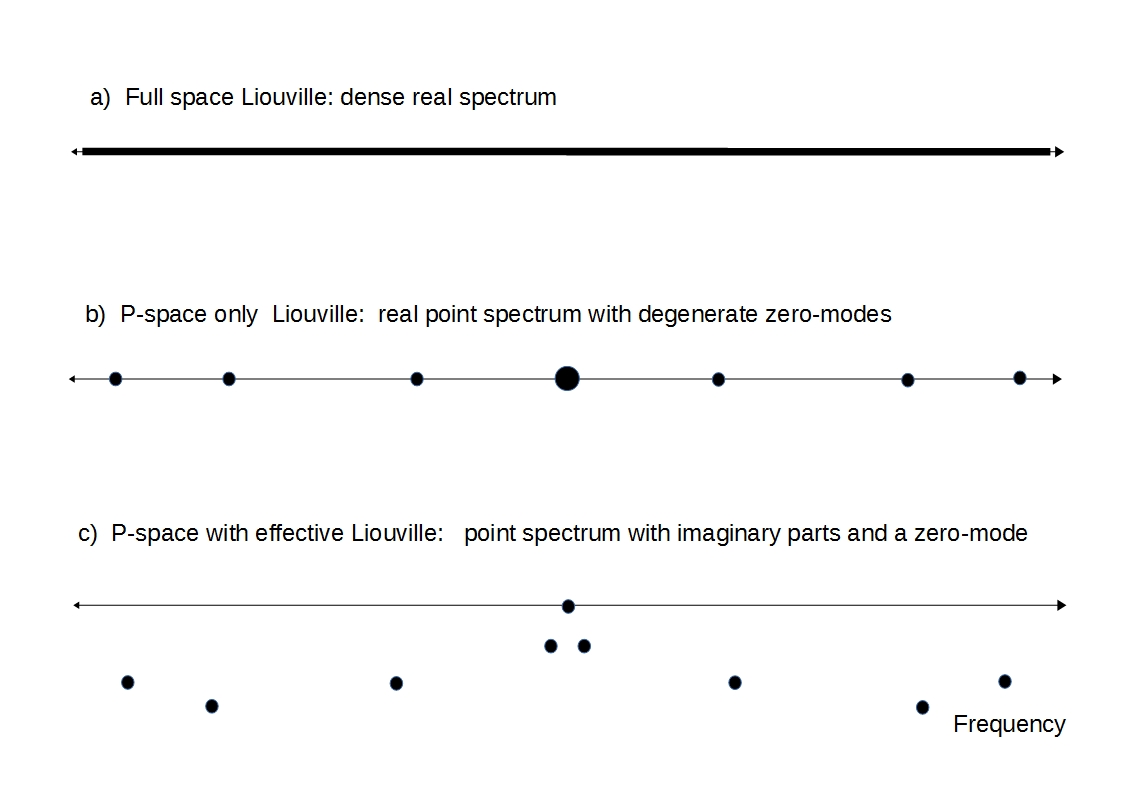}
\caption{Sketch of spectral properties of the total Liouville, the closed system's Liouville and of the  open system's effective Liouville}
\label{Fig5.1}
\end{figure}

While the form of Eq.~(\ref{last}) reflects the normalization of the density operator at each value of $t$ (  $\Tr \rho_\infty=1$, $\rl{k}$ are traceless) the hermiticity of $\rho$ is not obvious. This can be made obvious by  adding its hermitian conjugate and taking half of both,
\be
            \fbox{$\displaystyle \rho(t) = \rho_\infty + \sum_{k=1}^N    \frac{1}{2} \LK  A_{k} {e^{-i\omega_kt}} +  A^{\dagger}_{k}{e^{i\omega_kt}} \RK e^{-\gamma_kt}$}
	 \, , \label{lastt}
\ee
where the traceless matrices $A_k=a_k\Lambda_k$. The hermiticity requires that effective eigenvalues appear in pairs ($\omega_k-i\gamma_k$ and $-\omega_k-i\gamma_k$)  for finite $\omega_k$, or alone ($-i\gamma_k$) for vanishing $\omega_k$.  For example, if $N$ is odd, there must be at least one other eigenvalue $-i\gamma_k$  with vanishing frequency $\omega_k$ .

For some observable $\hat{O}=\hat{O}^\dagger$ the time evolution of its expectation value $O(t)=\Tr \hat{O} \rho(t) = (\hat{O} \mid \rho(t))$ reads accordingly
\be
            O(t) = O_\infty + \sum_{k=1}^N       |{a_O}_{k}|  \cos\LK \omega_k t + {\phi_O}_k \RK e^{-\gamma_kt}
	 \, , \label{lasttO}
\ee
where ${a_0}_k=(\hat{O}|A_k)=  |{a_O}_{k}|e^{-i{\phi_O}_k}$ are complex strengths from the overlap of $\hat{O}$ with matrix $A_k$ and $O_\infty$ is the stationary value of observable $\hat{O}$.

For an open quantum system with an environmental spectrum being relative smooth with respect to the system's spectrum  Eq.~(\ref{lastt}) yields the full time evolution of the reduced density  operator (up to additional   terms of type $t^{n}e^{-iz_kt}$ in case of accidental degeneracies). Note that we did not rely on any kind of perturbation theory nor on a Markov approximation. The non-Markovian character shows up in the positions of the effective eigenvalues $z_k=\lambda_k(z_k)$  and in the appearance of the derivative $\lambda_k'$. A Markov approximation would treat the effective eigenvalues as $z-$independent eigenvalues of the effective Liouville. Thanks to the Laplace transform the spectral representation is shown to be of the same form with and without a Markov approximation - only the correct position of (effective) eigenvalues and the amplitudes $a_k$ will differ.

Due to the exponential damping factors $e^{-\gamma_k t}$ in Eq.~(\ref{lastt}) the density operator will finally relax  to $\rho_\infty$, which is the zero-mode of the effective Liouville and  does not depend on the initial state $\rho_0$. The initial state enters the expression (\ref{last}) only via the prefactors $a_k$ of traceless eigenmatrices and which decay  exponentially in $\rho_t$.  Accordingly (see Eq.~(\ref{lasttO})) expectation values of observables  relax  to their stationary values,  showing a superposition of exponentially damped oscillations. The time scale of final relaxation  is set by the smallest value of $\gamma_k$ for which ${a_O}_k$  is non-vanishing. This substantiates statement 2. of the introduction.

We now like to give few remarks on the zero-mode and arguments in favor of statement 3. of the introduction. As shown above, the existence of a zero-mode is a consequence of the conservation of probability ($\sp{1}{\rho(t)}=1=\Tr \rho(t)$). As the coupling to an environment introduces the non-hermitian term to the effective Liouville, the high degeneracy  of the $0$-eigenvalue of the closed system (a consequence of unitary time evolution - symmetry with respect to time translation) will be lifted on opening the system and generically there will be no symmetry left which could introduce systematic degeneracies in a finite dimensional secular equation. Other degeneracies will be accidental and are unstable against any small change in the environment. Thus, for a generic opening (no symmetry left) the zero mode will be non-degenerate, i.e. unique after normalization. There is, however, one obvious exception when a system separates into two independent subsystems. In such case there will exist two zero-modes, one for each subsystem.  We have not tried to find a more formal proof for the uniqueness of the zero-mode, but  similar findings with more formal arguments  along a variant of the Perron-Frobenius theorem and numerical evidence for randomly chosen quantum maps are presented in \cite{BruzdaEtAL, BruzdaEtAL2}. In these works the spectrum of the quantum map  is addressed directly and not the  spectrum of the effective Liouville. They are however related by taking $\Phi_k=e^{iz_k}$ as the spectrum for the quantum map. This spectrum is restricted to the unit disc and the zero-mode corresponds to $\Phi_0=1$ (invariant state) and all other eigenvalues have moduli less than $1$  
($\gamma_k > 0$).

From Eq.~(\ref{lastt}) we can  also conclude most of the remaining   introductory statements, as will be substantiated in the following sections.

To put Eq.~(\ref{lastt}) in a broader context we discuss briefly  its limitations. If the environment has singularities in its spectrum (gaps, edges, isolated points) Eq.~(\ref{lastt}) will not be correct, unless one makes by hand a Markov approximation ($\DL$ independent of $z$), which enforces isolated simple poles in the resolvent $\LBK z-{\cal L}\RBK^{-1}$ for a finite dimensional $\cal L$.  In non-Markovian cases of structured environment the Laplace back transform cannot rely on simple poles and has to be calculated accordingly. Typically, the long time behavior will still depend significantly on the initial state, as not all non-zero modes of the Liouville will lead to exponential damping.  Such situations are of importance if one likes to use a system as a quantum probe  of properties of the environment (for examples see \cite{Xiong} and Chap.~IV of \cite{BreuerEtal}) or in applications of quantum information processing, where a structureless environment is disadvantageous, as it absorbs information from the system.  In the following, however, we stick to the situation of a structureless environment and study how then the system, described by Eq.~(\ref{lastt}),  evolves  to stationarity in the presence of such structureless environment.
\section{Time Evolution of Density Matrix Elements}
To discuss the time evolution of density matrix elements one has to choose a basis. In the absence of certain symmetries the only preferred  choice in hindsight  is the eigenbasis of the stationary state (zero-mode), as this state emerges as the characterizing timeless  state of the system.  However, one is free to choose some other basis before one knows the stationary state and we consider the density matrix in the matrix representation of eigenstates $\LMID n\KET$ of some seemingly relevant observable
$\hat{O}=\sum_n o_n \LMID n \KET \BRA n \RMID$. 
\be
           \rho_{mn}(t) = {\rho_\infty}_{mn} + \sum_{k=1}^N   \frac{1}{2} \LK  {A_{k}}_{mn} {e^{-i\omega_kt}} +  {A^{\dagger}_{k}}_{mn} {e^{i\omega_kt}} \RK e^{-\gamma_kt } \, .  \label{last2}
\ee
The hermiticity and trace preserving properties of the density matrix are obvious in the representation of Eq.~(\ref{last2}). To guarantee  also the positivity of the density matrix, the matrix elements ${A_{k}}_{mn}$  and  the eigenvalues $=\omega_k-i\gamma_k$ have to fulfill constraints. Such constraints will be fulfilled automatically if these values are derived without approximations  from the constructed effective Liouville. If, instead, they are used as modeling parameters, these constraints have to be respected in the modeling.

If a certain left eigenstate has, by accident,  a vanishing trace with the initial density matrix,  then ($A_k=0$)  it will not contribute to any matrix element of $\rho_{nm}(t)$. The density matrix elements evolve as a superposition of damped irregular oscillations, since each eigenvalue $z_k=\omega_k-i\gamma_k$ contributes with some amplitude and oscillation (for finite $\omega_k$), damped by the exponential factor $e^{-\gamma_k t}$. Let us order those with $A_k \not=0$  in increasing order, $\gamma_1 < \gamma_2 < \gamma_3 \, \ldots $. The shortest value  $\gamma_1$  sets the longest relaxation time $\tau=(\gamma_1)^{-1}$ of modes contributing to the density matrix. Every non-vanishing matrix element of the corresponding right eigenmatrix, denoted as $\hat{\Lambda}$,  will affect the relaxation of  matrix elements of the density matrix to their stationary value. 

Now, we are able to substantiate statement 4. of the introduction.  In the matrix representation of the stationary states eigenbasis  the time scale $\tau$ sets the common scale of  relaxation of diagonal elements (called populations)  and of the decay  of off-diagonal elements (called coherences) provided   the eigenmatrix $\hat{\Lambda}$ (corresponding to $\tau$) in this representation has non-vanishing matrix elements at the  matrix positions considered. 
It may happen that $\hat{\Lambda}$  has a number of vanishing matrix elements. However, it is not to be expected in a given basis (eigenbasis of $\hat{O}$ or eigenbasis of $\rho_\infty$) that the traceless eigenmatrices turn out to be sparse with many vanishing entries, as the only requirement is that they form a complete bi-orthonormal set with their left partners.

Statement 5. of the introduction doesn't need a special argument in this context as it is always the case that two hermitian matrices (here $\hat{O}$ and $\rho_\infty$) can be diagonalized together if and only if they commute. To reach an approximate common diagonalization their commutator has to be quantified as small, e.g. by comparison of its Hilbert-Schmidt norm with the Hilbert-Schmidt norm of the two ways of building their product.

To elucidate statement 6. of the introduction we look at the often discussed situation where all diagonal elements of all traceless  right eigenstates vanish. This very special situation means that the diagonal elements of the density matrix (populations) are stationary from the very beginning and not influenced by the opening. It requires a very subtle coupling of the system to the environment to reach such situation. It means that the relevant observable $\hat{O}$ is still a constant of motion, as its expectation value will not change in time.  In such situation only off-diagonal elements change in time and reach their stationary values in an exponential damped way with characteristic time $\tau$. If in addition, the relevant operator $\hat{O}$ commutes to a high degree with the stationary density matrix  (quantitatively controlled by the ratio of the Hilbert-Schmidt norm of the commutator with that of the products alone), this very special situation corresponds to so-called pure dephasing where all off-diagonal elements (called coherences) decay to zero. The characteristic time scale of this pure dephasing  is set by $\tau$. Since a re-population is excluded by an exact symmetry there is no characteristic time scale for relaxation of populations (it is often treated as infinite). 

Let us come back to generic  situations  with  some relaxation of populations and substantiate statement 7. of the introduction. 
Since every non-vanishing matrix element of $\hat{\Lambda}$ will affect the relaxation of  matrix elements of the density matrix to their stationary value, this time scale is characteristic for all of these matrix elements. It may happen that all  diagonal matrix elements vanish. Then another higher value of $\gamma_k$ will set the time scale for populations, resulting in slightly faster re-population as the decay to stationarity of those coherences affected by 
$\Lambda$. However, $\tau$ and this next shorter time scale will not differ by orders of magnitude.  Also the opposite special situation may occur, where $\Lambda$ is purely diagonal and $\tau$ sets the scale of relaxation of populations and $\gamma_2$ sets the somewhat  smaller characteristic time scale for decay of many coherences to their stationary value. Also in this special case the time scales are of comparable magnitude.
Only in very  special situations each eigenmatrix $\Lambda_k$ may be very sparse and have only two or few finite entries.  For such situation to occur the coupling between system and environment  has to designed in a sophisticated  way adapted to the relevant operator $\hat{O}$, or to its zero mode $\rho_\infty$. Then, each matrix element of the density matrix  will represent  a single eigenmatrix or very few eigenmatrices  with individual decay rates. Only in such  sophisticatedly designed situations  one could  have separated time scales of decay for different matrix elements and  a separation of time scales for slow  relaxation (set by $\tau$)  of populations and for fast decay of some coherences (set by  very large values of $\gamma_k$). 
In a  generic case of system-environment coupling, however, Eq.~(\ref{last2}) tells that  $\Lambda$ will effect diagonal and off-diagonal matrix elements of the density matrix and exponential decay of coherences to their stationary value and relaxation of populations   to their stationary  value will be characterized by a common time scale $\tau$.
In the eigenbasis of the stationary state the generic situation will have the populations relax to their stationary value on a common time scale $\tau$ with coherences decaying to zero. The vanishing of off-diagonal matrix elements in the eigenbasis of  a  relevant operator $\hat{O}$ can only be precise to that extent as its commutator with the stationary density operator can be neglected.

As a typical example -consistent with \Eq{last2}-  of a density matrix time evolution for a two-value system take the following modeling with decay rates $\gamma_1$ and $\gamma_2$ and frequency $\omega$, two real prefactors $r$ and $s$, a real phase $\phi_0$ and the stationary population $p_\infty$ of the first value. $\Lambda_1$ with four non-vanishing matrix elements and $\Lambda_2=(\Lambda_3)^\dagger$ with three non-vanishing matrix elements and  bi-orthonormal left-partners where chosen in a basis where $\rho_\infty$ is diagonal,
\bea
\rho_{11}(t)&=&p_\infty +s e^{-\gamma_1 t}  +2 r \cos(\omega t+\phi_0) e^{-\gamma_2 t}=1-\rho_{22}(t)\, ,\nonumber\\
\rho_{12}(t)&=& s e^{-\gamma_1 t}  +i r e^{-i\omega t-i\phi_0-\gamma_2 t}=\rho_{21}^\ast(t) \, .
\eea
The parameters have to fulfill the constraints
\be
0\leq \rho_{11}(t) \leq 1 \; ; \;\det \rho(t) \geq 0 \, .
\ee
In this modeling $\gamma_1$ and $\gamma_2$ are not necessarily ordered, such that the smaller decay rate of both  sets the scale $\tau$ of relaxation to the stationary state and the decay of coherences.

To make contact with the widespread folklore that decay of coherences typically happens much faster than relaxation of populations to stationarity we come to the conclusion of statement 8. of the introduction. A separation of time scales for population relaxation and decoherence  involves two different couplings of the system to two different types of environment in order to be consistent with Eq.~(\ref{last2}) in a  generic situation (not designed for special purposes). A plausible way is to think of a fast relaxation process with short $\tau_1$ by contact to an almost elastic interaction with many scattering modes. Here coherences may decay very quickly to an intermediate stationary value (it would be stationary in the absence of the second environment) and populations will be redistributed only slightly, but on the same time scale.  An additional inelastic equilibrating contact with the second environment  with much slower but perhaps more substantial relaxation of the populations (and the remaining perhaps tiny coherences)  then leads to the final stationary state. 
\section{Time Evolution of the Entropy}
We will look at the entropy of the system's relevant density operator as it captures essential differences between reversible and irreversible dynamics. The entropy quantifies the information about the spreading of actual states over all possible basic states.
For a system with $d$ basic states it can very between $0$ (1 basic state with probability one) and $\ln d$ (all basic states with equal probability $1/d$). In classical  Hamiltonian dynamics the states will be  redistributed without change of the occupied phase space volume (Liouville's theorem) and  the entropy stays constant. This expresses the reversible character of classical Hamiltonian physics.  In  standard (non-quantum)  Markov stochastic processes the so-called relative entropy (actual relative to  stationary)  is a Lyapunov function (a non-negative function with non-positive derivative which vanishes with vanishing slope at the stable stationary point of a dynamics) (see e.g~Chap.~V.5. in ~\cite{VKam}). This Lyapunov function property  of relative entropy highlights the irreversible character of standard Markov processes. In quantum stochastic process of closed systems, oscillations and a constant entropy highlight the reversible character of closed quantum processes. Therefore, we like to see what happens to the entropy in processes of open quantum systems.

The entropy of the system's relevant density operator $\rho(t)$ is defined in the usual way,
\be
 S(t)=\Tr \LB \hat{S}(t) \rho(t)\RB=\sp{\hat{S}(t)}{\rho(t)}\, ,
\ee
where $\hat{S}:=-\log \rho$ is the entropy operator (see~\cite{ZubEal}). For later use we mention that the conservation of probability $\Tr \rho(t) =1$ and the fact that $\rho(t)$  and $\hat{S}(t)$ commute
yields (in scalar product notation)
\be
\sp{\dot{\hat{S}}(t)}{\rho(t)}=0\, . \label{ent2}
\ee
With this relation, the time derivative of the entropy reads
\be
\dot{S}(t)=\sp{\hat{S}(t)}{\dot{\rho}(t)}\, .\label{ent1}
\ee

Let us recall the invariant character of entropy in   a closed quantum  system, where the time evolution is unitary.
Since $\hat{S}(t)$ is a function of $\rho(t)$ both transform in the same way under the unitary time evolution $U(t)$ and we
end up with
\be
	S(t)=\Tr \LB U(t) \hat{S}(t=0) U^\dagger(t)  U(t) \rho(t=0) U^\dagger(t)\RB = S(0) \, ,
\ee
due to the unitary invariance and cyclicity of the trace. In a closed system the reversible quantum dynamics does not change the information about the spreading of the states over all possible basic states; a satisfying result.
It also possible to see this by invoking Eq.~(\ref{ent1}) and using the von Neuman equation, the commutator structure being reminiscent of the unitary time evolution.

It is clear that the information about the spreading of states within the open quantum system will not be a constant of motion any longer. 

Firstly, we may directly  write down the expectation value of $S(t)$ with the help of Eq.~(\ref{lastt}) and arrive at an equation which looks similar to the expectation value for an observable (Eq.~(\ref{lasttO})),
\be
	  S(t) = S_\infty + \sum_{k=1}^N       |{a_S}_{k}|(t)   \cos\LK \omega_k t + {\phi_S}_k(t)  \RK e^{-\gamma_kt}
	 \, , \label{lasttS}
\ee
where ${a_S}_k(t) =(\hat{S}(t)|A_k)=  |{a_S}_{k}|(t) e^{-i{\phi_S}_k(t)}$ depend on time. The time dependence of these coefficients  is a crucial difference to the case of an observable which is time independent (we have adopted the Schr\"odinger picture throughout). In this time dependence the $\log$ of the time dependent density operator enters again and Eq.~(\ref{lasttS}) does not offer the explicit time dependence of the entropy.
Nebvertheless, we can see that it will approach a stationary value $S_\infty$ and other contributions die out exponentially fast. We can also see that the largest deviation  from $S_\infty$ after time $\tau$  will come from the $\hat{\Lambda}$ mode,
\be
	  S(t\gtrsim \tau) = S_\infty +        |{a_S}|(t)   \cos\LK \Omega t + {\phi_S} (t)  \RK e^{-t/\tau}
	 \, , \label{lasttSS}
\ee
where $a_S(t)$ is the complex amplitude of that mode and $\Omega$ is its frequency.  For $\Omega=0$ (overdamped case) (or $\Omega \tau \ll 1$ in an underdamped case) one can neglect (or approximately neglect) the oscillating behavior and the entropy approximately approaches its stationary value in a monotonic way after time $\tau$. When $\omega$ is not much smaller than $1/\tau$, the entropy still oscillates until it reaches its stationary value.  

Now, we will investigate the non-negative relative entropy (see e.g.~\cite{Schu,Wehrl}) of the time dependent state with respect to the stationary state,
\be
	  S_{\rho(t),\rho_\infty} :=\Tr \rho(t)\log\rho(t) -\Tr \rho(t) \log \rho_\infty  \geq 0
	 \,  \label{lasttSS1}
\ee
In scalar product notation it reads   $S_{\rho(t),\rho_\infty}=\sp{\hat{S}_\infty - {\hat{S}}(t)}{\rho(t)}$.
One can define another relative entropy which is non-negative as well by
 \be 
	S_{\rho_\infty,\rho(t)}:=\sp{\hat{S}(t) - {\hat{S}}_\infty}{\rho_\infty} \geq 0\, .
\ee
We now calculate the production  of the first relative entropy,  use Eq.~(\ref{ent2}), and find
\be
	 \dot{ S}_{\rho(t),\rho_\infty} =\sp{\hat{S}_\infty - {\hat{S}}(t)}{\dot{\rho}(t)} 
	 \, . \label{lasttSS2}
\ee
We know the density operator has irregular oscillations for times well below the relaxation scale $\tau$. Therefore we consider now times $t\gtrsim \tau$. In this regime the density operator can be approximated as
\be
             \rho(t\gtrsim \tau) = \rho_\infty +  \frac{1}{2} \LK  \hat{A} {e^{-i\Omega t}} +  \hat{A}^{\dagger} {e^{i\Omega t}} \RK e^{-t/\tau}
	 \, . \label{ent4}
\ee
In an overdamped case (or in an underdamped case with very low frequency $\Omega \tau \ll 1$), we can approximate
\be
             \dot{\rho}(t\gtrsim \tau) = - \frac{1}{\tau} \LK \rho(t\gtrsim \tau) - \rho_\infty\RK  
	 \, . \label{ent5}
\ee
Consequently the relative entropy production reads in this approximation
\be
	 \dot{ S}_{\rho(t),\rho_\infty} =\frac{1}{\tau}\sp{\hat{S}(t) - {\hat{S}}_\infty}{{\rho}(t)} 
- \frac{1}{\tau}\sp{\hat{S}(t) - {\hat{S}}_\infty}{{\rho}_\infty} \leq 0
	 \, , \label{lasttSS3}
\ee
and thus can serve as a Lyapunov function for the stability of the stationary state. This substantiates the first part of statement 9. in the introduction. From the foregoing discussion the second part is also justified, as in all other cases the non-positivity of the relative entropy production is not valid due to regular or irregular oscillating contributions.

\section{Conclusions}
Our substantial  conclusions are already  listed as 9  statements in the introduction.

The present comment could have been written more than 50 years ago and the author wonders if he has overlooked such contribution when studying the literature.

{\bf Acknowledgements: }
I thank J\'anos Hajdu for bringing the work \cite{JPol_statdyn}  of Janos Polonyi to my attention and for initiating discussions between the three of us about time scales in open quantum systems.  
The present comment  is  an attempt to address  some of the questions raised during these discussion with my toolbox of generated dynamics. I also thank Rochus Klesse for clarifying discussions. The responsibility for any misconception in this comment  is with the author.

\end{document}